\documentclass{elsart}

 \usepackage{subfigure}
 \usepackage{graphics}
 \usepackage{graphicx}
\usepackage{lscape} 
\usepackage{rotating}
\usepackage{amssymb}

\begin{document}

\begin{frontmatter}



\title{An Investigation into the $^{113}$Cd Beta Decay Spectrum using a CdZnTe Array}

 \author[apc]{J. V. Dawson,}
\author[sussex]{C. Reeve,}
\author[oxford] {J. R. Wilson,}
\author[dresden]{K. Zuber,}
\author[lngs]{M. Junker,}
 \author[dortmund]{C. G\"{o}ssling, T. K\"{o}ttig, D. M\"{u}nstermann, S. Rajek, O. Schulz}
\address[apc]{Laboratoire Astroparticule et Cosmologie, 10 rue Alice Domon et L\'eonie Duquet, 75205 Paris, France}
\address[sussex]{University of Sussex, Falmer, Brighton. BN1 9QH, UK}
\address[oxford]{University of Oxford, Denys Wilkinson Building, Keble Road, Oxford. OX1 3RH, UK}
\address[dresden]{Institut f\"{u}r Kern- und Teilchenphysik, Technische Universit\"{a}t Dresden, Zellescher Weg 19, 01069 Dresden, Germany}
\address[lngs] {Laboratori Nazionali del Gran Sasso,  Assergi, Italy}
\address[dortmund]{Lehrstuhl f\"{u}r Experimentelle Physik IV, Technische Universit\"{a}t Dortmund, Otto-Hahn Str. 4, 44227 Dortmund, Germany}

\author{}

\address{}

\begin{abstract}%
We present 11 independent measurements of the half-life and spectral
shape of the 4-fold forbidden beta decay of
$^{113}$Cd using CdZnTe semiconductors with a total combined lifetime of 6.58 kg days. Our overall result gives a half-life of $(8.00 \pm 0.11(stat) \pm 0.24(sys))\times10^{15}$ years and a Q value of $322.2 \pm 0.3(stat) \pm 0.9(sys)$ keV. For the first time half-lives well beyond 10$^{10}$ years have been deduced 
with a statistically representative sample of independent measurements.
\end{abstract}

\end{frontmatter}

\section{Introduction}
\label{intro}
In the history of particle and nuclear physics the study of weak interactions and especially beta decay has played a vital role. These studies helped, amongst others, to establish the V-A structure of weak
interactions. Nowadays, this interest is somewhat reduced but there
are still interesting topics to investigate like the endpoint measurements of tritium and $^{187}$Re electron spectra to determine the neutrino mass \cite{bon08} or the search for S, T, V contributions to the weak interaction \cite{bec06}.
In addition to these beyond the standard model searches, some interesting nuclear physics questions are 
still open like the study of highly forbidden beta decays. The major bulk of beta emissions are characterised
as allowed or single forbidden, however there are a few isotopes which are at least 4-fold forbidden, having ft-values beyond 20 \cite{sin98}. These are extremely rare decays with half-lives well beyond
$10^{10}$ years. It occurs that for 5-fold forbidden transitions (like in $^{48}$Ca and $^{96}$Zr) even double beta decay is more likely to occur.\\
In this paper the focus is on 4-fold forbidden ${\beta}$-decays.
There are only three nuclei in nature which permit a feasible study of
four-fold forbidden beta decay, $^{113}$Cd, $^{50}$V and $^{115}$In,
all of them are non-unique ($\Delta I ^{\Delta \pi} = 4^+$). Half-lives of these transitions are long ($\geq10^{14}$ years)
and in typical experiments would produce very low count rates, and as
such can only be studied in well shielded, low radioactive background
experiments.  In this paper the focus is on the decay $^{113}$Cd
(1/2$^+) \rightarrow ^{113}$In ($9/2^+$) , with a Q-value of 320 $\pm$
3 keV \cite{wapstra}.\\ 
The COBRA collaboration is performing a search for neutrinoless double
beta decay of which the half-life may be well above $10^{21}$ years
\cite{zuber2001}.  Results from previous COBRA experiments can be
found in \cite{Kiel2003,Munstermann2003,Bloxham2007}.  The present
COBRA experiment, known as the 64-array \cite{Munstermann2007}, is
formed of 64 1 cm$^{3}$ CdZnTe semiconductor crystals, each with a
mass of $\sim$ 6.5g.  The experiment is shielded and situated
underground in Laboratori Nationali del Gran Sasso (LNGS) in
Italy. Due to the cadmium content of the semi-conductor, this
four-fold forbidden non-unique beta decay of $^{113}$Cd forms the
dominant low energy feature.  We present results from the first layer of this experiment comprising 16
crystals. Measurements of the half-life of this decay and Q-value have
been made independently for 11 working and well-behaved crystals.\\
There have been five previous attempts to measure the half-life of $^{113}$Cd. The first using a CdTe device
resulted in (4-12)$\times$ 10$^{15}$ years \cite{Mitchell_1998}.  Measurements using CdWO$_{4}$ as a scintillator
found 7.7$\pm$0.3$\times$10$^{15}$ years \cite{Danevich_1996} and as a
cryogenic bolometer obtained
9.0$\pm$0.5(stat)$\pm$1(sys)$\times$10$^{15}$ years
\cite{Alessandrello_1994}.  Previously the COBRA collaboration
obtained 8.2$\pm$0.2(stat)$^{+0.2}_{-0.1}$(sys)$\times$10$^{15}$
years \cite{Kiel2005,Goessling_2005} using room temperature CdZnTe
semiconductor detectors. A very recent result using a CdWO$_{4}$
scintillator measured 8.04$\pm$0.05$\times$10$^{15}$ years
\cite{Belli_2007}.  We present spectra from 11 detectors, with a total
exposure of 6.58 kg days.
\section{Experimental Setup}
\label{setup}
Measurements were performed with the first installed layer of the
COBRA 64-array, comprising sixteen 1 cm$^{3}$ CdZnTe semiconductor
detectors.  The array is housed in an inner copper shield of 5 cm
thickness, and surrounded by a lead castle of 20 cm thickness.\\
The whole setup is enclosed in a Faraday cage, which itself is surrounded by a neutron 
shield made out of 7 cm thick boron-loaded polyethylene plates and 20 cm of paraffin wax. The full
experiment is situated in Laboratori Nationali del Gran Sasso (LNGS) with 1400 m of rock shielding, corresponding to about 3500 metres water equivalent.\\
\begin{figure}
\begin{center}
  \includegraphics[width=5cm]{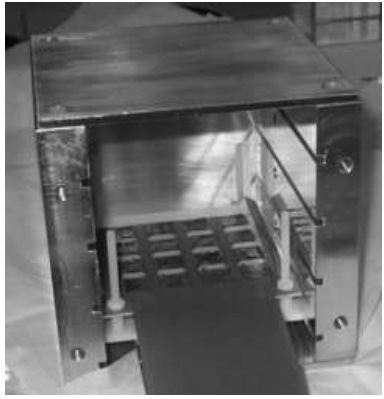}
\caption{\small First 16 crystal layer inside copper crystal holder
  during commissioning.}
\label{fig:backgroundmodel}
\end{center}
\end{figure}
The CdZnTe detectors were manufactured by eV-PRODUCTS
\cite{EVPRODUCTS}, and are CoPlanar Grid (CPG) devices\cite{luke1995} with gold anode grids and cathode. Detectors of this kind only read out the electron signal and their operational principle is analogous to a Frisch grid in a wire chamber. The detectors are coated with a passivation 
paint to provide a highly resistive surface layer and to maintain
longterm stability.  The detectors are bonded to kapton signal and
high voltage foils with a low activity copper loaded glue which was
developed in-house.  Each crystal receives individually tailored high voltage, grid bias and weighting of the CPG
subtraction circuit.  These parameters are optimised to minimise the
Full Width Half Maximum of the 1274.5 keV photopeak from a $^{22}$Na
calibration source.  The CPG guard rings are not
contacted to maximise the active mass of each crystal, typically 6.5g. Each crystal has a slightly different mass, which are known having being weighed by the manufacturers prior to
passivation; Table \ref{tab:Resolutions-of-detectors2}.\\
All the COBRA electronics have been designed and manufactured in-house.
The CPG anode signals are processed by a 16-channel preamplifier based
on the suggested circuit from eV-PRODUCTS.  The resulting signals are
shaped by COBRA shaping amplifiers with time constants of 1$\mu$s,
digitised by COBRA four-channel 14-bit peak sensing ADCs which are
integrated into VME architecture. The manufacturer (Analog Devices) specifications of
the 14-bit peak sensing ADC chip, AD7865 \cite{AD7865}, give a maximum DNL (Differential
Non-Linearity) of $\pm$1LSB (Least Significant Bit).  All ADCs have a
slight variation on the width of each ADC code (voltage range which
results in a particular ADC code) and this can be observed in spectra
with high numbers of counts, such that the Poisson uncertainty is
insignificant in comparison with the variation of each ADC code width.
The resulting effect on a spectrum is that neighbouring ADC values
appear to be statistically incompatible with each other.  These
effects are not confused for real spectral features since the former
occur on a bin-by-bin basis, and spectral features are usually far
wider.  In some situations this effect can be considered as a
systematic or indeed calibrated for.  If, as in this case, spectra
from different ADCs or with different calibrations as with a long
running experiment are summed energy-wise then the effect becomes
statistical.  It was found that an assumption of an uncertainty of
10\% on each ADC code was enough to bring neighbouring values into
statistical agreement for the whole COBRA dataset.  All data-sets
presented, therefore, have statistical uncertainties that include a 10\%
uncertainty on the bin-width.  For low statistics data-sets, such as for
double-beta decay studies, the number of counts are too low to observe this effect and only the study of the $^{113}$Cd spectrum is affected.\\
The COBRA setup is regularly calibrated with $^{22}$Na (511 and 1274.5
keV photopeaks),$^{57}$Co (122.1 keV photopeak) and $^{228}$Th (2614.5
keV photopeak) sources.  A linear relationship between energy and ADC
channels is found.  Independent confirmations of the external
calibration are made by measuring photopeaks present in the combined
low background spectrum.  Two prominent photopeaks are observed in the
low background data, at 351.2$\pm$0.9 keV and 606$\pm$2 keV, these are
identified as the 351.9 keV and a blend of 609.3 keV and 583 keV lines
from the $^{238}$U and $^{232}$Th chains.  We take the uncertainty on the
351.9 keV line, 0.9 keV, as a representation of the systematic uncertainty
of the energy calibration of the entire experiment. This peak is close
to the $^{113}$Cd shoulder and is therefore a very relevant
calibration point.  It was observed through-out the entire duration
of the experiment and so if any calibration drifts did occur it would
be subject to the same bias as the data.  The uncertainty on the peak
position is slightly greater than the bin-width of the spectra, 0.7
keV, and is probably dominated by the number of counts observed, and
therefore can be considered as a conservative estimate.\\ 
The relationship between the energy resolution (quantifiable in terms of Full Width Half Maximum) and photopeak
energy is found to closely follow the functional form $FWHM=\sqrt(a^2 + (b.E)^2)$, where a and b are fit parameters.  Table \ref{tab:Resolutions-of-detectors2}
shows these resolution equations determined for the operational detectors.
Runs were taken in units of 1 hr.  Hour-long runs were deemed abnormal and rejected from the analysis if the number of counts observed in that run did not conform to the overall observed mean count rate.  If the probability of the count rate for each hour-long run being compatible with the mean count rate was lower than 1\% then that run was rejected from the analysis.  Such runs on closer inspection showed rapid bursts of events, thought to be due to vibrations, micro-discharges on the high voltage cables and electronic effects.%
\begin{table}[H]
\begin{centering}
\caption[Detector
  Properties.]{\label{tab:Resolutions-of-detectors2}Detector
  properties: mass of crystals, and FWHM resolution equations in keV with E, energy, in keV.}
\begin{tabular}{ccc}
Detector & Mass(g) &Resolution Equation \tabularnewline
\hline
\hline
{1} &{6.483}   & {none}          \\[-1.5ex]
{2} &{6.454} &{ $\sqrt( (13.261 \pm 0.196)^2 + ((0.02843 \pm 0.0003)E)^2)   $   }     \\[-1.5ex]
{3} &{6.454}  &{$\sqrt( (10.616 \pm 0.301)^2 + ((0.02714\pm 0.0003) E)^2)   $   }     \\[-1.5ex]
{4} &{6.524}   & {none}          \\[-1.5ex]
{5} &{6.512}  &{$\sqrt( (10.037 \pm 0.239)^2 + ((0.02373 \pm 0.0002)  E)^2)   $   }    \\[-1.5ex]
{6} &{6.459}  &{none} \\[-1.5ex]
{7} &{6.526}   & {$\sqrt( (20.649 \pm 0.156 )^2 + ((0.02964 \pm 0.0003 ) E)^2)   $    }    \\[-1.5ex]
{8} &{6.465}  &{$\sqrt( (13.937 \pm 0.636 )^2 + ((0.04718 \pm 0.0006 )E)^2)   $  }   \\[-1.5ex]
{9} &{6.469} &{ $\sqrt( (10.502 \pm 0.254 )^2 + ((0.02638 \pm 0.0003 )E)^2)   $    }   \\[-1.5ex]
{10} &{6.461} &{ $\sqrt( (48.588 \pm 2.950 )^2 + ((0.07473 \pm 0.0041 )E)^2)   $      }        \\[-1.5ex]
{11} &{6.468} &{$\sqrt( (14.673 \pm 0.303 )^2 + ((0.03786 \pm 0.0004  )E)^2)   $ }    \\[-1.5ex]
{12} &{6.492} &{$\sqrt( (11.570 \pm 0.317 )^2 + ((0.02639 \pm 0.0004  )E)^2)   $   }   \\[-1.5ex]
{13} &{6.529} &{$\sqrt( (33.630 \pm 1.824 )^2 + ((0.05804 \pm 0.0023 )E)^2)   $  }   \\[-1.5ex]
{14} &{6.548} &{$\sqrt( (11.672 \pm 0.243 )^2 + ((0.02384 \pm 0.0003  )E)^2)   $ }    \\[-1.5ex]
{15} &{6.520} &{$\sqrt( (10.703 \pm 0.395 )^2 + ((0.02771 \pm 0.0005 )E)^2)   $ }  \\[-1.5ex]
{16} &{6.509}  & {none}                          \\[-0.5ex]
\end{tabular}
\end{centering}
\end{table} 

\section{Background Model}
\label{background}
As common in rare event searches a background model can be constructed. This is based on the measurement of radioactive contaminants in material samples. This typically includes the natural
decay chains of U and Th, $^{40}$K and potential isotopes produced by cosmic ray spallation while the materials were on surface. With the given activities a theoretical spectrum can be constructed which should agree with  the observable spectrum if all impurities are included. To account for detector effects, 
the background spectrum for the experiment is simulated using a GEANT 4 based simulation code.  This simulation includes the 16 crystals with passivation paint,
delrin crystal holder, copper and lead shielding. \\
In our case, the detector components have been assessed by the LNGS low background counting facility. Table \ref{table:contamination} shows the overall results. Only the passivation paint which coats 5 sides of each crystal has definitive measurements, all the rest are upper limits.  This passivation paint is anticipated to be a major background contributor to the experiment.  \\
\begin{table}[pth]
\caption{List of Contaminants}
{\footnotesize
\begin{tabular}{@{}crrr@{}}
{Material} &{Isotope} &{Activity (g/g)}   \tabularnewline
\hline
\hline
{Copper} &{Th-232} &{$<5.7\times10^{-10}$}   \\[-1.5ex]
{} &{U-238} &{$<2.0\times 10^{-10}$}   \\[-1.5ex]
{} &{U-235} &{$<2.5\times 10^{-9}$}    \\[-1.5ex]
{} &{K-40} &{$<3.6\times 10^{-6}$}    \\
\hline   
{CdZnTe} &{Th-232} &{$<12\times 10^{-9}$}   \\[-1.5ex]
{} &{U-238} &{$<41\times 10^{-10}$}   \\[-1.5ex]
{} &{U-235} &{$<9\times 10^{-9}$}    \\[-1.5ex]
{} &{K-40} &{$<8.4\times 10^{-6}$}    \\
\hline   
{Delrin} &{Th-232} &{$<12\times 10^{-10}$}   \\[-1.5ex]
{} &{U-238} &{$<4\times 10^{-10}$}   \\[-1.5ex]
{} &{U-235} &{$<4\times 10^{-9}$}    \\[-1.5ex]
{} &{K-40} &{$<10\times 10^{-7}$}    \\
\hline   
{Passivation Paint} &{Th-232} &{$(2.7\pm0.2)\times 10^{-7}$}   \\[-1.5ex]
{} &{U-238} &{$(1.7\pm0.1)\times 10^{-7}$}   \\[-1.5ex]
{} &{U-235} &{$(3.0\pm0.5)\times 10^{-7}$}    \\[-1.5ex]
{} &{K-40} &{$(2.2\pm0.3)\times 10^{-4}$}    \\
\end{tabular}\label{table:contamination} }   
\end{table}
 Figure \ref{fig:backgroundmodel} shows the summed
background spectrum from all 11 crystals (higher rate spectrum with error bars).  The dominant low energy feature (below $\sim$320 keV) is the beta
decay spectrum of $^{113}$Cd, a theoretical spectrum is shown (dot dashes). Superimposed on top of this are
background spectra arising from contaminants in the local
environment.\\
The main identified contaminants are radon
gas, and trace uranium and thorium found in the crystal passivation
coating.  Simulations of these
contaminants have been made and are shown in Figure~\ref{fig:backgroundmodel} (small squares and dashes).\\
To address this issue new crystals with a colourless
passivation coating were purchased and briefly operated in the shielded
setup.  This new coating was tested in the LNGS low counting facility,
and is far cleaner than the one coating the 16-crystals, with only upper
limits found, $<$34 ppm $^{40}$K, $<$11 ppb $^{238}$U and $<$45 ppb $^{232}$Th.  The spectrum
obtained with one colourless crystal is shown in Figure
\ref{fig:backgroundmodel} (square data points with error bars).\\
Of equal concern is the presence of radon gas inside
the inner detector volume.  The experiment has recently been moved
from the outer parts of LNGS laboratory to a new location between 
the main halls A and B (which is
known to be less abundant in radon), and as a result the overall
background count rate has fallen.  Brief experimentation with flushing
with nitrogen boil-off gas from a liquid nitrogen dewar and filtering
with a radon filter has shown improvements in background
rates, supporting the hypothesis that radon gas is observable in the
background spectrum.  The spectrum from the new colourless crystals
showed far less of a reduction in background than anticipated from the
contamination measurements. Their spectrum is assumed to be dominantly
contaminated by radon. In effect these measurements give the radon
contamination levels. Simulated $^{222}$Rn in the air cavity of the
inner detector volume is shown in Figure \ref{fig:backgroundmodel} (dashes) as well, and is scaled to fit the colourless data.\\
Finally the sum of these components; paint contamination and radon, and
the theoretical $^{113}$Cd spectrum, is shown in Figure
\ref{fig:backgroundmodel} (thin continuous line).\\
\begin{figure}
\begin{center}
  \includegraphics[width=15cm]{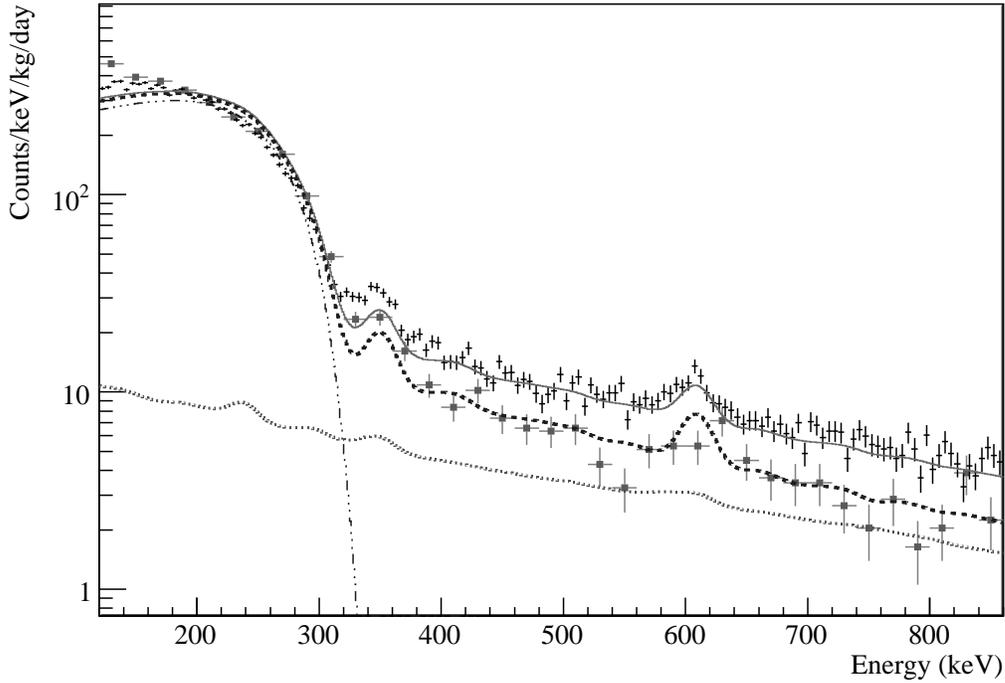}
\caption{\small Summed spectrum from all detectors (higher rate spectrum with error bars),
  compared to crystal with new passivation coating (square marker with error bars). Overlaid
  are the simulated backgrounds produced by the passivation coating (light dashes) and theoretical $^{113}$Cd beta spectrum (dot-dashes).  The effect of radon gas and combined with the  $^{113}$Cd beta spectrum (bold dashes), overlaps the data from the new crystals.  The sum of these components (thin continuous line) overlaps the spectrum from all detectors. The 351.9 keV and 609.3 keV $\gamma$-lines from the $^{238}$U decay chain are visible in the spectrum.}
\label{fig:backgroundmodel}
\end{center}
\end{figure}
The simulated paint spectrum is blended with the energy resolution of each detector and the resulting spectrum is subtracted from each real data spectrum.  The radon levels are unknown and may also change over time and so could be different for each detector.  However, as can
be seen in Figure \ref{fig:backgroundmodel} the Rn level is basically given by the measured spectrum of
the colourless painted detector and has a small effect on the total rate of the $^{113}$Cd spectrum
as described in more detail later. The simulated radon spectrum is blended with the energy resolution of each detector and the resultant spectrum is fitted to the real data spectrum with the range 350 to 700 keV.
If the background is known and the simulation accurately represents the experiment, then each real data spectrum should consist only of the contribution from the $^{113}$Cd spectrum.  These 11 spectra are then used to determine the Q value and half-life of the $^{113}$Cd spectrum.\\

\section{Sources of Systematic Uncertainties}
\label{Systematics}
The following subsections describe the main sources of systematic uncertainties.  Table \ref{table:systematics} shows the comparison between the differing effects. The dominant sources of systematic uncertainty are the cadmium content and the energy calibration.  The uncertainties are asymmetric since the deadlayer contribution can only reduce the crystal mass. We assume that these uncertainties are independent and therefore add them quadratically to obtain the total uncertainty of $+2.9,-3$\%. As we have many detectors we can attempt to estimate the systematic uncertainty, see Section \ref{Discussion}. \\
In the collaboration's previous attempt to measure the $^{113}$Cd
half-life \cite{Goessling_2005}, the largest source of uncertainty
came from the uncertainty in the detector deadlayer. This has now been
improved and it can be shown that this systematic was previously
heavily over-estimated.\\
   \begin{table}[pth]
\caption{Table of Systematics}
{\footnotesize
\begin{tabular}{@{}cc@{}}
{Origin} & {Fractional contribution to half-life} \tabularnewline
\hline
\hline
{Cd content} & {$\pm 0.02$}  \\
{Deadlayer} & {$-0.009$} \\
{Crystal masses} & {$\pm0.00015$} \\
{Detection Efficiency} & {$\pm0.0002$}  \\
{Background Subtraction} &{$\pm 0.0058$}\\ 
{Energy calibration} & {$\pm 0.02$}\\
\hline  
{Total (quadratic)} & {+0.029, -0.03}\\
\end{tabular}\label{table:systematics} }
\vspace*{-13pt}   
\end{table}

\subsection{Cadmium Content and crystal mass}
\label{cadmium_proportion}
CdTe semiconductors are well known to suffer from polarisation
problems. This issue was addressed with the addition of zinc, such
that CZT detectors are of the form Cd$_{1-x}$Zn$_{x}$Te with a
proportion of zinc, {\it x}, typically of 0.1. Due to the production
method, boules of material are grown and the crystals cut from this.
The zinc quantity can be non-uniform throughout the entire boule such
that detectors cut from one part may not have the same zinc content as
crystals cut from another, and hence there exists a systematic
uncertainty on the corresponding quantity of cadmium in each crystal. 
Based on advice from
the manufacturers, the zinc admixture was assumed to be between
7-11\%, such that the proportion of cadmium could be between 89-93\%.
Since all the detectors are different and could have come from different locations in the boule we therefore assume an average value of 91\% to calculate the half-life for each crystal, and include a systematic uncertainty of 2\% in the half-life.\\
The mass measurements per crystal, as given by the manufacturers, are shown in Table \ref{tab:Resolutions-of-detectors2} and are assumed to have an uncertainty of 0.001 g, or 0.015\%
\subsection{Crystal Deadlayer}
It is possible that the crystals possess an inactive layer of CdZnTe,
thus reducing the effective mass of the detectors. Observations of
alpha signals from an $^{214}$Am source with peak alpha energy of 5.5
MeV penetrating all sides of a painted CZT crystal indicate that if
deadlayers exist they are of the order of $\sim$15$\mu$m, and can not
be substantially larger. The presence of a deadlayer of the order of 15$\mu$m would reduce the amount of active mass of the detectors by 0.9$\%$.
\subsection{Detection Efficiency}
\label{efficiency}
As the CdZnTe detectors are small (1 cm$^{3}$) some proportion of the
$^{113}$Cd betas will escape the detector without fully depositing
their energy, slightly distorting the observed spectrum.  Complete
escapes are also possible. Using the COBRA simulation tool $^{113}$Cd
betas were simulated uniformly throughout a crystal volume and the
resulting spectrum compared to the theoretical spectrum used.  For a
threshold of 120 keV, the probability of observing a $^{113}$Cd beta
compared to the theoretical spectrum is found to be 0.9932. So we can
anticipate that 0.68\% of the betas escape or deposit less energy than
120 keV in each detector.  This efficiency must be factored into the
total half-life estimate. The uncertainty on this quantity is assumed
to be 0.0002.
\section{Background Subtraction} %
The background is assumed to be due to known quantities as discussed
earlier in Section \ref{background}.  The background level is
anticipated to be slightly different for each crystal and each data
set as radon levels fluctuate with time.  The systematic uncertainty on the half-life from this component is calculated from the uncertainty on the background model fit to each spectrum.  Table \ref{table:counts_subtracted} details the amount subtracted from the spectrum, assumed to be background, and the residual spectrum which is assumed
 to be a pure $^{113}$Cd spectrum.  Some $\sim$10\% of the data are
 removed from each spectrum in this process within the range 120 to 320 keV. Typically the uncertainty on the background fit is $\sim$6\%. Therefore the systematic uncertainty on the number of counts
 in the pure $^{113}$Cd spectrum is estimated to be 0.58\%.

  \begin{table}[pth]
\caption{Table of Background Counts Subtracted}
{\footnotesize
\begin{tabular}{@{}ccccc@{}}
{} &\multicolumn{4}{c}{Total Counts   counts/(kg.day)}    \\[-1.5ex]
{Detector} & {Background} & {Uncertainty} &{$^{113}$Cd} &{Uncertainty} \tabularnewline
\hline
\hline
{2}& {1537.4 }& { 86.1 }& {12630.3 }& {91.9 }    \\[-1.5ex]
{3}& {1257.8 }& { 76.7}& { 12166.9}& { 99.8 }    \\[-1.5ex]
{5}& {1718.8}& {  68.8}& { 13440.2 }& {94.4}     \\[-1.5ex]
{7}& {1446.0 }& { 72.3}& { 12497.1}& { 85.8}     \\[-1.5ex]
{8 }& {2122.8}& { 112.5}& {  12777.0 }& {138.9  }\\[-1.5ex]
{9}& {1392.0}& { 62.6 }& {13157.9 }& {111.5 }    \\[-1.5ex]
{10}& {1269.3}& { 100.3}& { 12018.5}& { 92.4 }   \\[-1.5ex]
{11}& {1102.5}& {49.6 }& { 11678.7}& { 108.6 }   \\[-1.5ex]
{12}& {1478.9}& { 63.6}& {  12670.8 }& {115.7 }\\[-1.5ex]  
{14}& {1463.6}& { 73.2}& { 12380.8}& { 85.2 }    \\[-1.5ex]
{15}& {1636.9}& { 65.5}& { 12980.3}& { 90.9  }   \\
\end{tabular}\label{table:counts_subtracted} }
\end{table}

\subsection{Energy Calibration}
The estimation of the systematic uncertainty based on the energy calibration
was made using the summed spectrum of all 10 detectors.  The observed
351.9 keV line, from the $^{238}$U chain, was fitted with a Gaussian
function returning a mean energy of 351.2 $\pm$0.9 keV.  This line was
used as it is the closest photopeak to the falling edge of the
$^{113}$Cd spectrum.  The falling edge of the beta spectrum highly
constrains the fit, so the energy calibration of this feature is the
most important.  We assume, therefore, that the energy calibration is
good to at least 0.9 keV.  The fit procedure described in the
following section was repeated with spectra shifted (both in the
positive and negative senses) by 0.9 keV.  The normalisation and hence
the half-life changed by $\pm$2\%.

\section{Fit Method}\label{fittingmethod}
The shape of the beta spectrum is theoretically described as 
\vspace{-10pt}\begin{equation}N(E) = F(Z,E).p.(E + m_e).(Q-E)^2 .S(E) \end{equation} \vspace{-20pt} where $F(Z,E)$, the Fermi function is interpolated from tabulated values found in \cite{bj69}. Here the electron momentum is $p(E)$, mass $m_e$ and kinetic energy $E$, and $Q$ is the endpoint of the beta decay spectrum.\\
Following the approach of \cite{Danevich_1996} the correction factor, $S(E)$, is assumed to be 
\vspace{-10pt}\begin{equation} S(E) = p(E)^6 + 7.c_1 .p(E)^4 .q(E)^2 + 7.c_2 .p(E)^2 .q(E)^4 + c_3 . q(E)^6 \end{equation}\vspace{-20pt} where $q(E)$ is the neutrino momentum. 
This form is for a four-fold unique beta decay, but has been found to fit the observed spectrum well by many authors  \cite{Kiel2003,Danevich_1996,Belli_2007}.\\
The energy resolution of the detectors smears the spectrum, but as the detector resolution changes with energy, it was necessary to use Monte Carlo techniques to simulate test spectra and perform chi-squared fits to the data. The assumed resolutions and functions are described in Section \ref{setup}. A MINUIT routine was devised, with the following free parameters
\begin{itemize}
\item Q - the Q value of the beta transition.
\item The spectral $S(E)$ parameters; $c_1$, $c_2$ and $c_3$.
\item The amplitude of the spectrum which is related to the half-life
\end{itemize}
To determine the half-life for each crystal the outputted fit parameters were used to determine the integral number of counts between 0 and the determined Q value.  As each parameter has an associated uncertainty, including the Q value, Gaussian random deviates were drawn to give estimates of each parameter and the resulting beta spectrum integrated. This procedure was repeated for each detector 1000 times, and the determined integral counts averaged. The standard deviation of this distribution represents the uncertainty on the integral due to the uncertainties on the fit.\\
The half-life was calculated according to: $T_{\frac{1}{2}} = \ln(2)
\eta_{det} N \frac{t}{S}$, where the efficiency $\eta_{det}$ is the
efficiency of detecting a beta decay, $N$ is the number of source
$^{113}$Cd atoms, $t$ is the measuring time and $S$ is the total number of
counts in the beta spectrum. Since we do not measure the entire beta
spectrum, energy thresholds are typically $\sim$100 keV, $S$ is the
integral of the fitted spectrum from 0 keV to the determined Q-value.
The efficiency $\eta_{det}$ accounts for the fact that some betas
above the energy threshold escaped, reducing the number of counts in
the observed spectrum.  This value  was taken as 0.9932 (as discussed in Section \ref{efficiency}).
The number of $^{113}$Cd source atoms is calculated for each detector, and assumes a natural abundance of $^{113}$Cd isotope of 12.22(12)\% \cite{chartofnuclides} and the overall proportion of cadmium with respect to zinc to be 0.91 (as discussed in Section \ref{cadmium_proportion}).

\section{Results}
\begin{figure}
\caption{$^{113}$Cd spectra and fit with c$_1$,c$_2$ and c$_3$ fixed.}
\label{fig:sub}
\subfigure[Detector 2] 
{
\includegraphics[width=6cm, height=3cm]{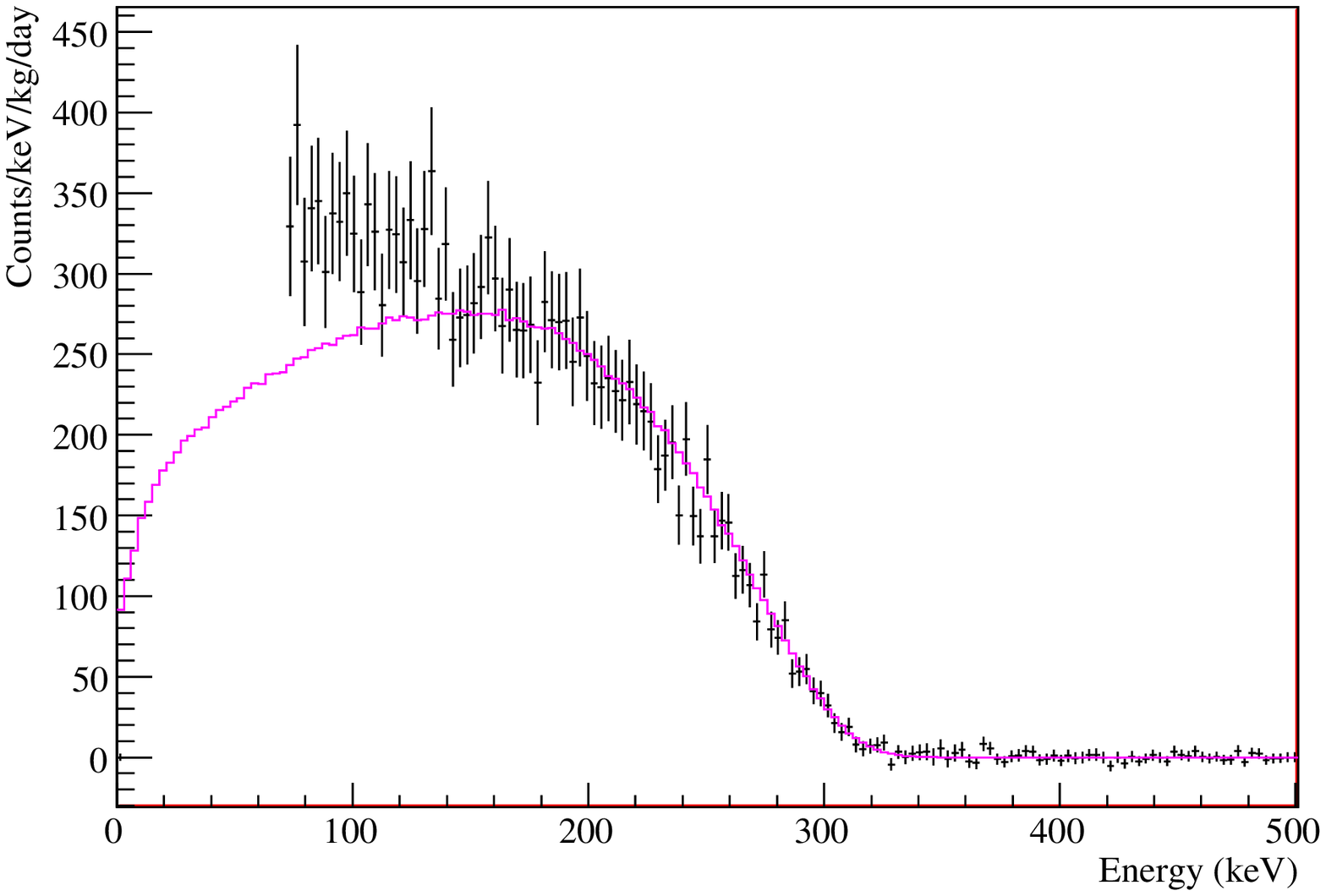}
}
\subfigure[Detector 3] 
{
\includegraphics[width=6cm,height=3cm]{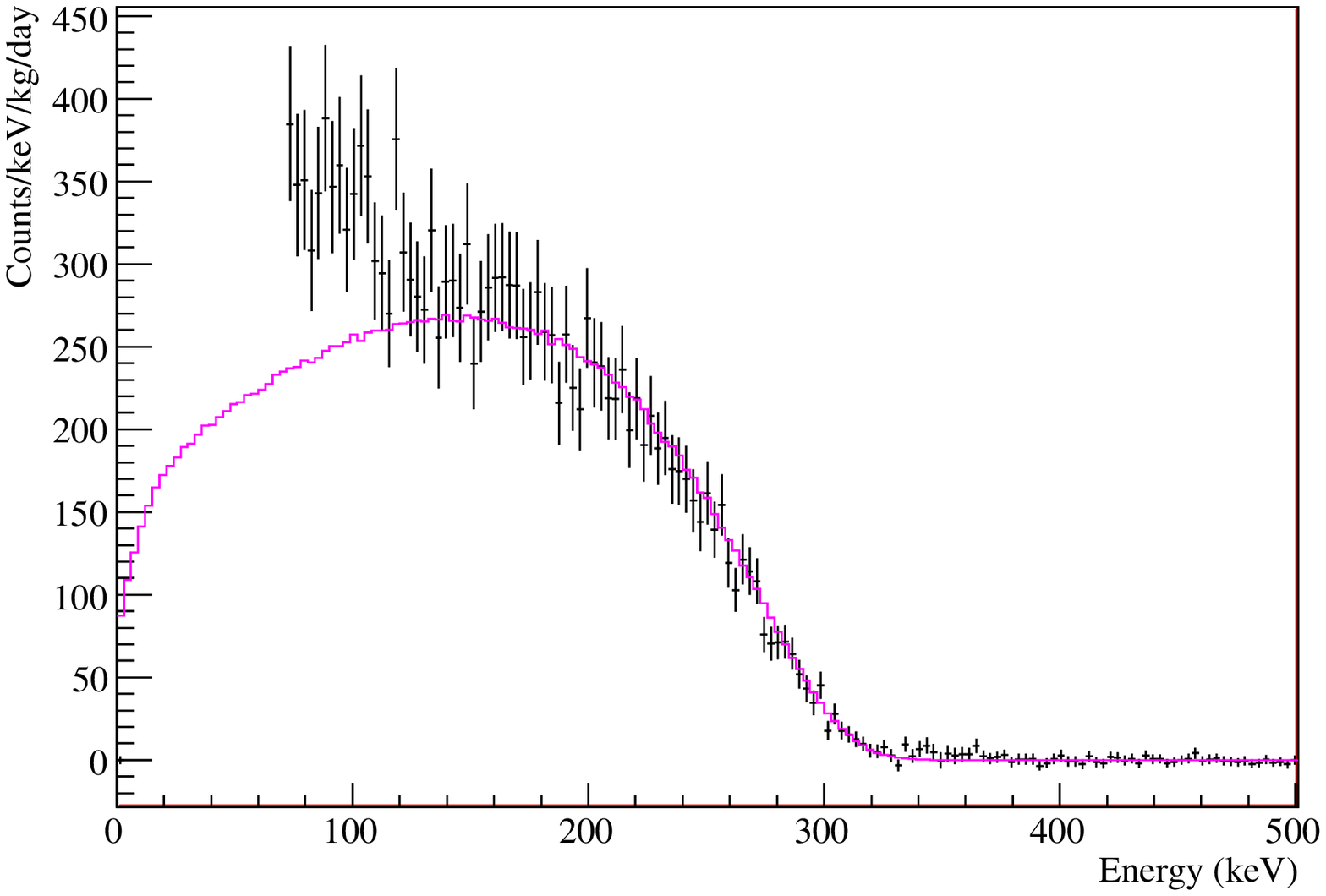}
}
\subfigure[Detector 5] 
{
\includegraphics[width=6cm,height=3cm]{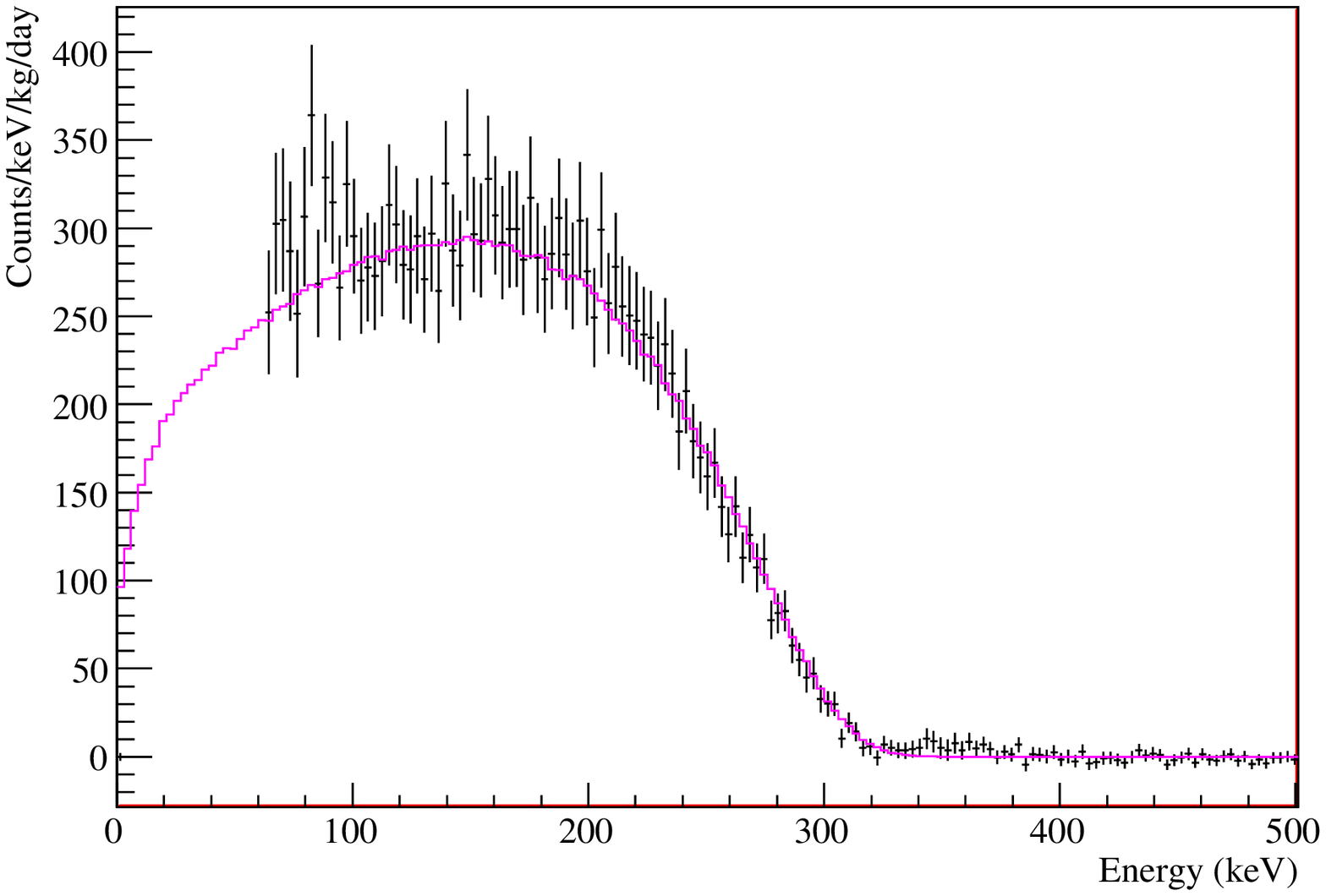}
}
\subfigure[Detector 7] 
{
\includegraphics[width=6cm,height=3cm]{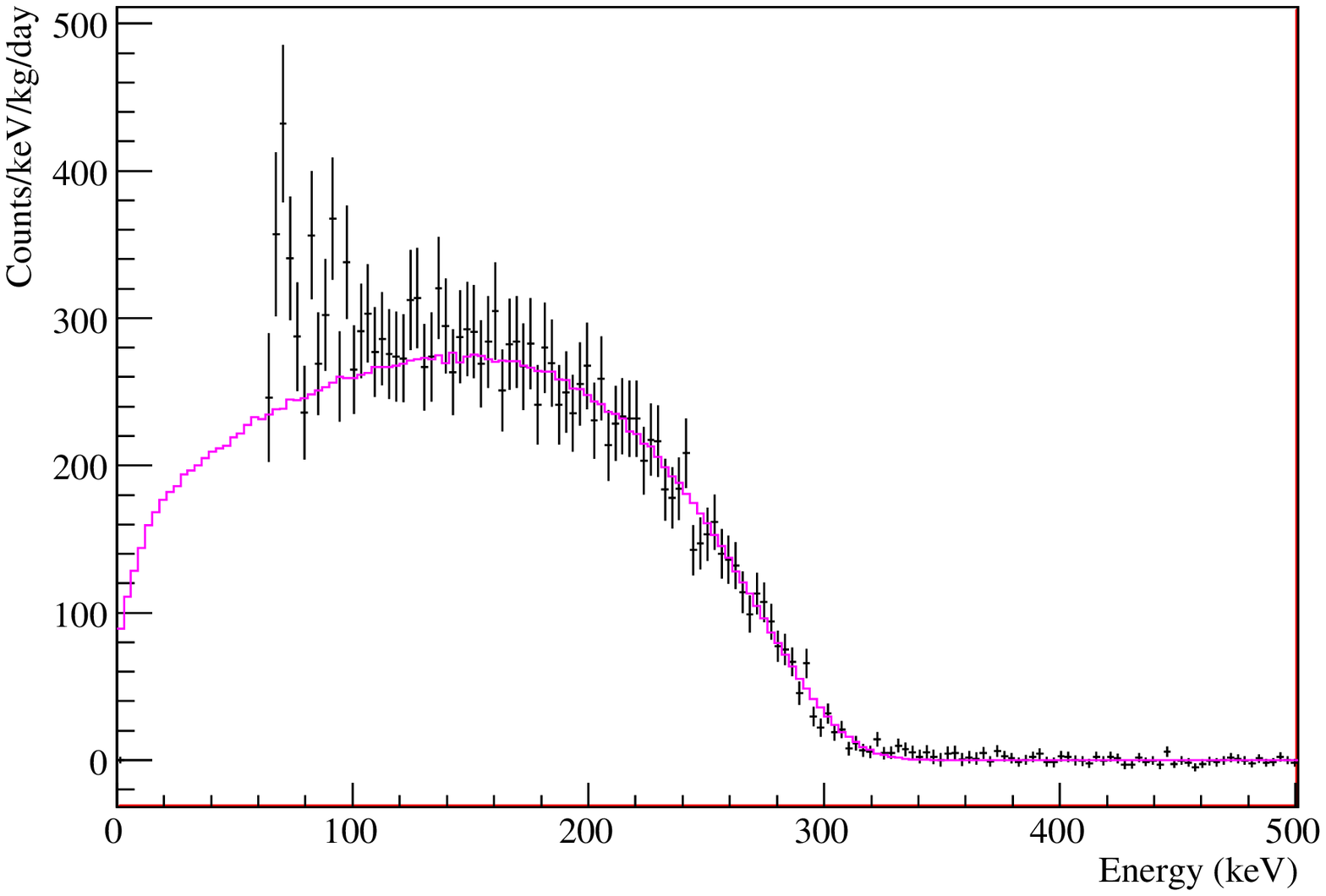}
}
\subfigure[Detector 8] 
{
\includegraphics[width=6cm,height=3cm]{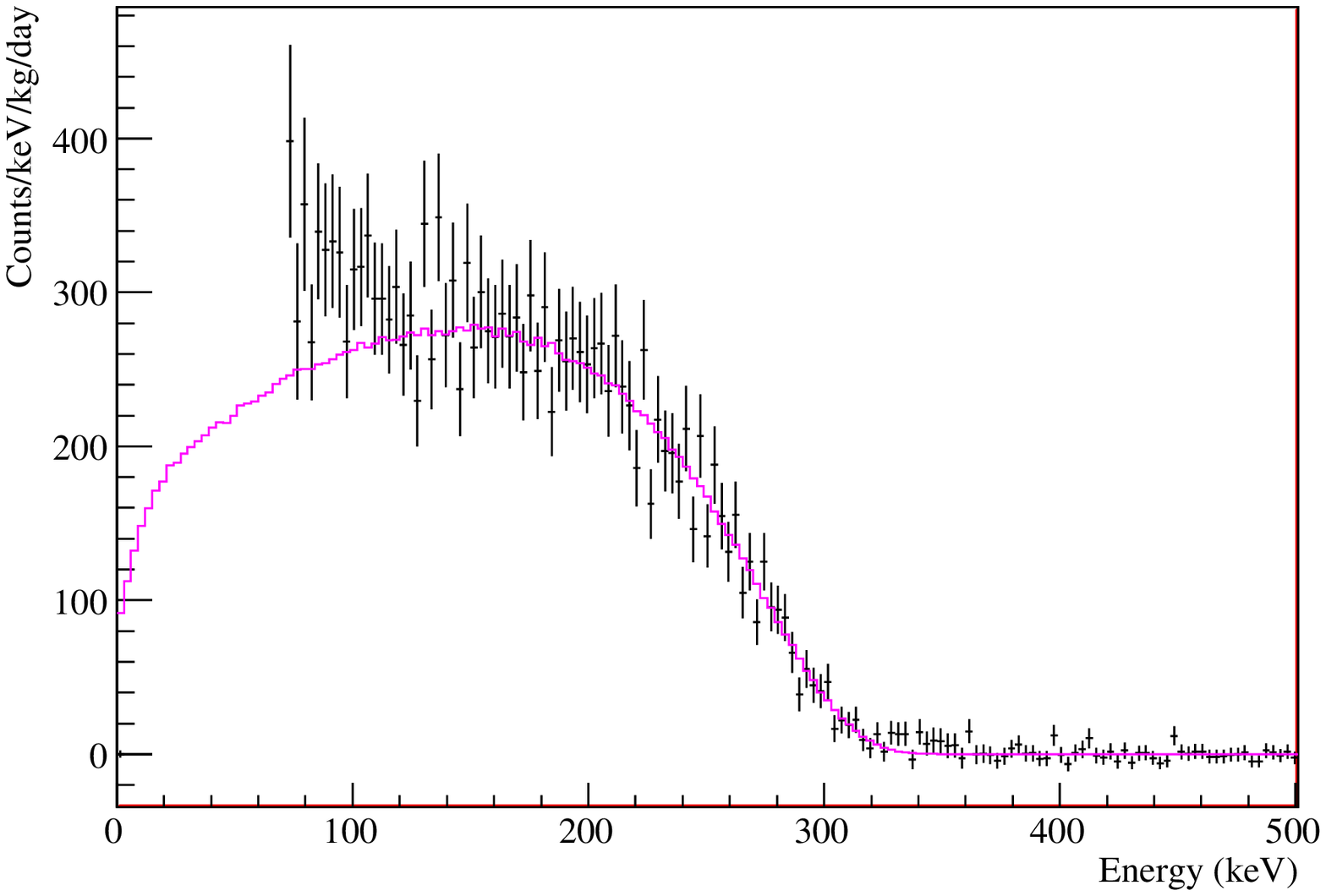}
}
\subfigure[Detector 9] 
{
\includegraphics[width=6cm,height=3cm]{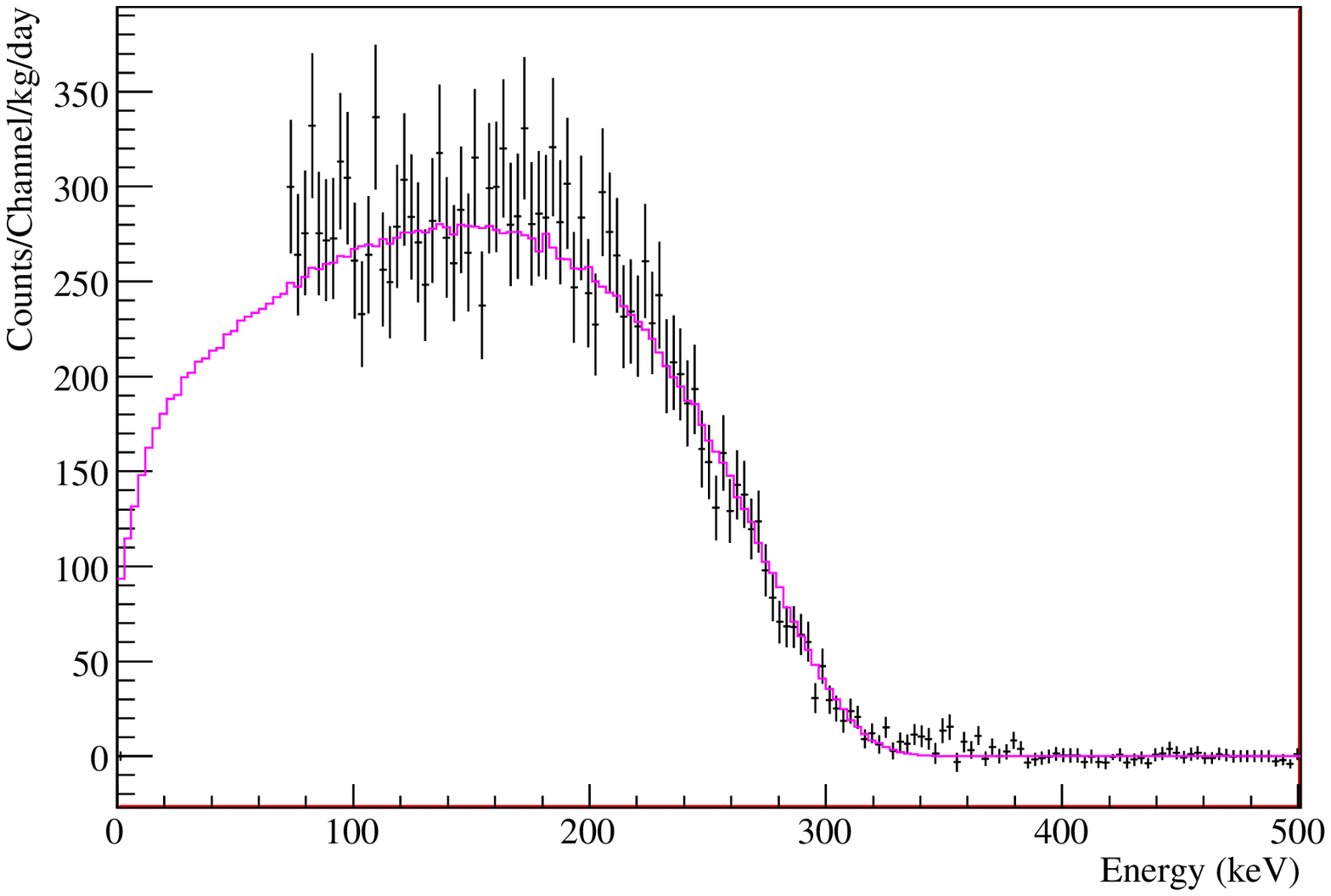}
}
\subfigure[Detector 10] 
{
\includegraphics[width=6cm,height=3cm]{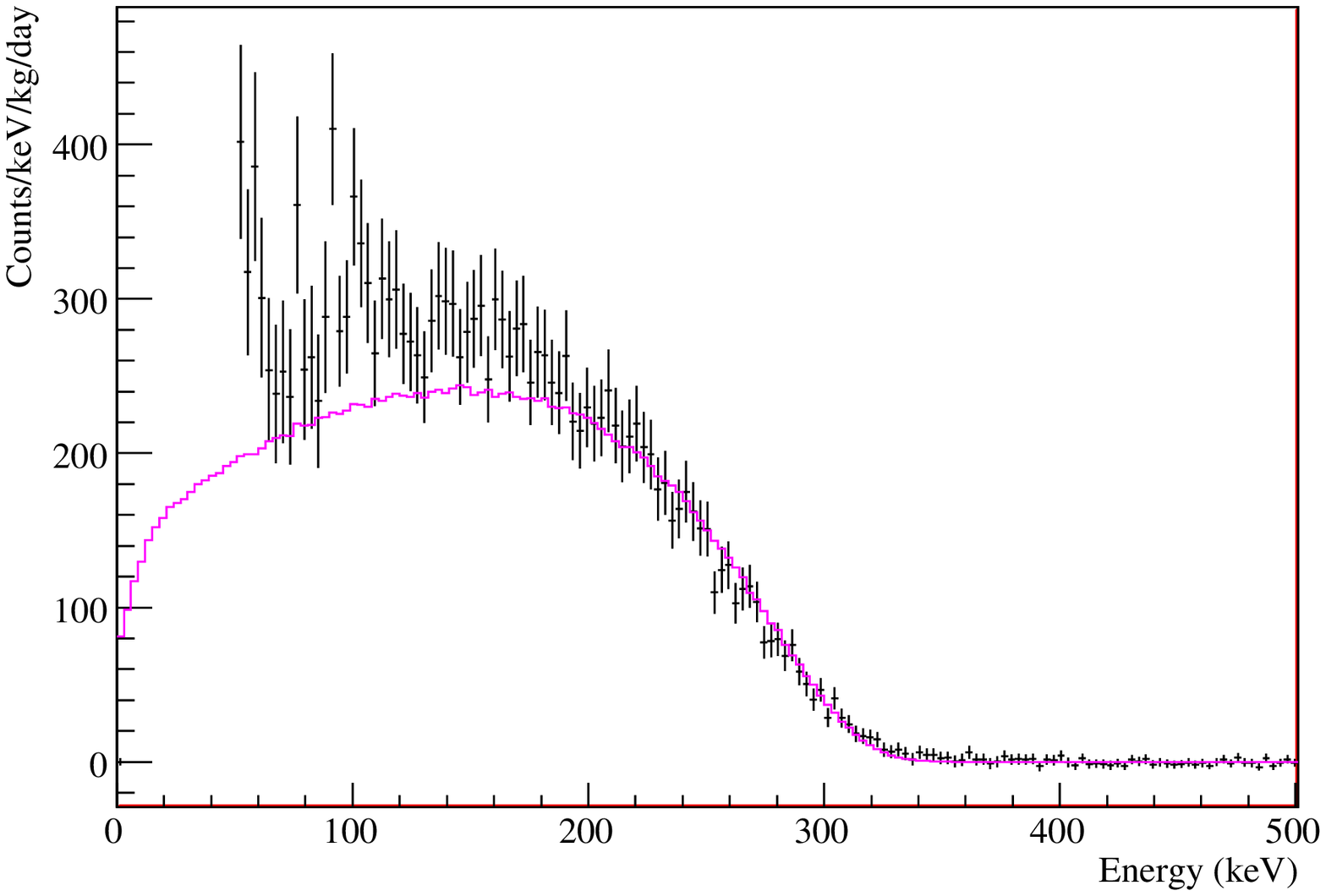}
}
\subfigure[Detector 11] 
{
\includegraphics[width=6cm,height=3cm]{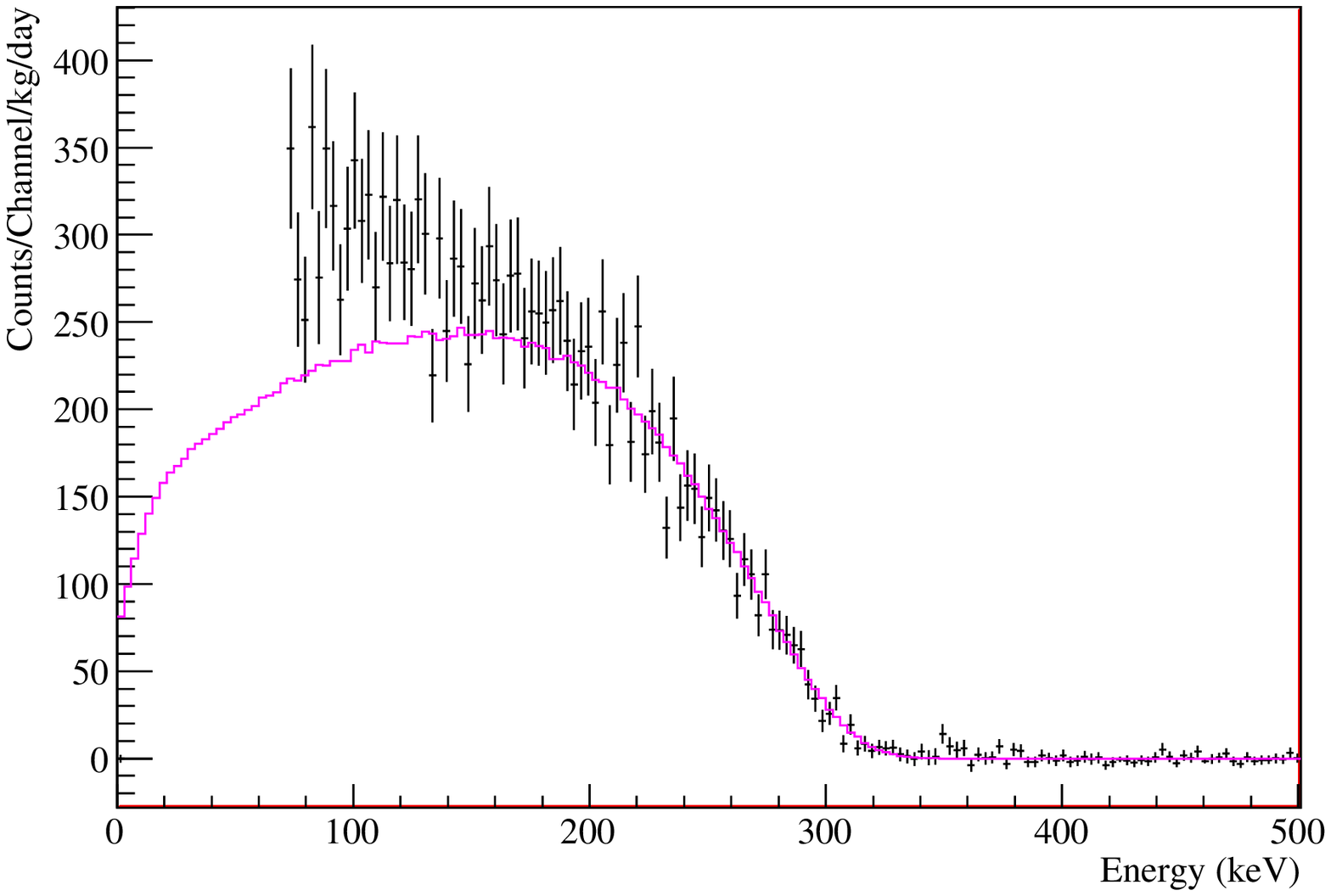}
}
\subfigure[Detector 12] 
{
\includegraphics[width=6cm,height=3cm]{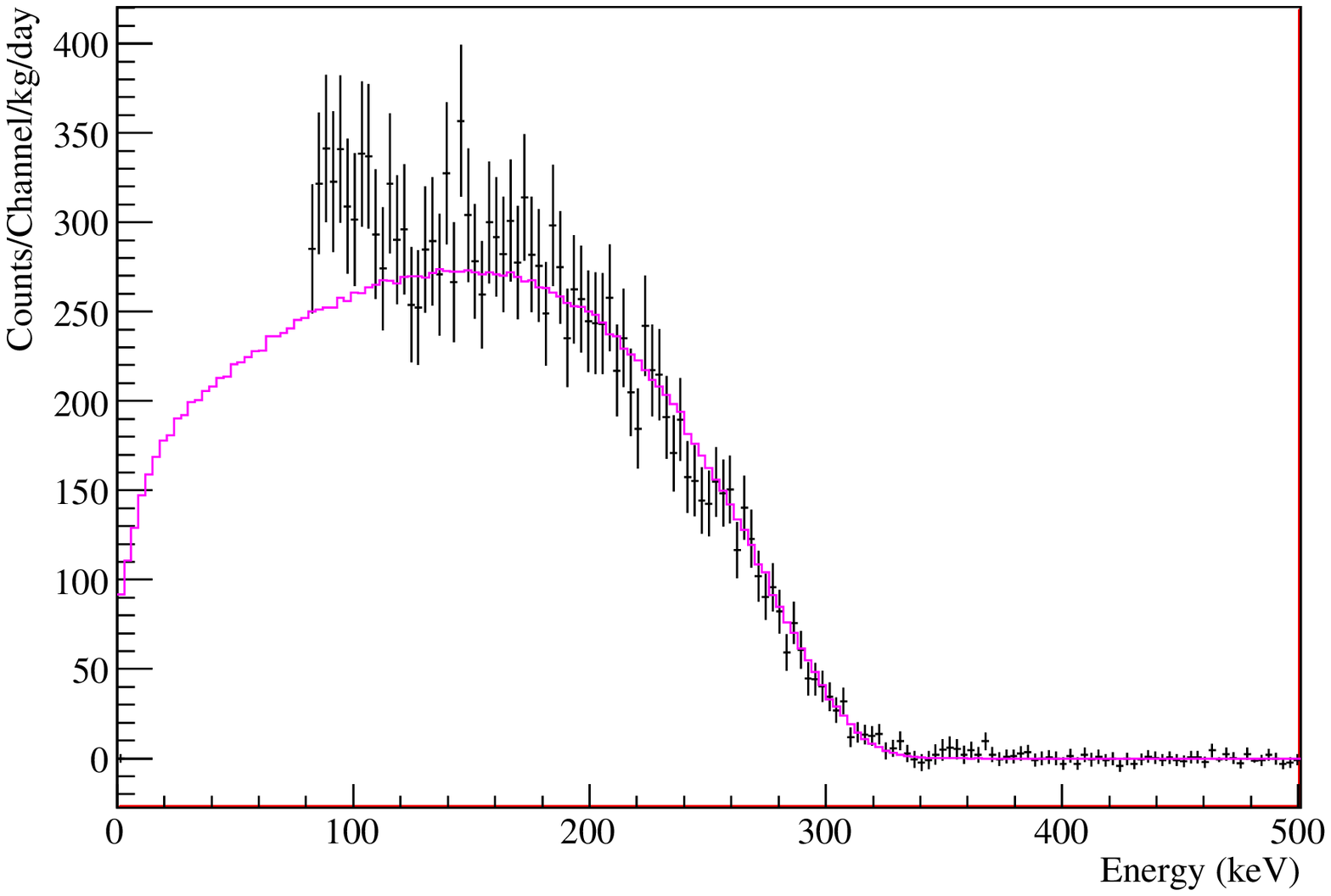}
}
\subfigure[Detector 14] 
{
\includegraphics[width=6cm,height=3cm]{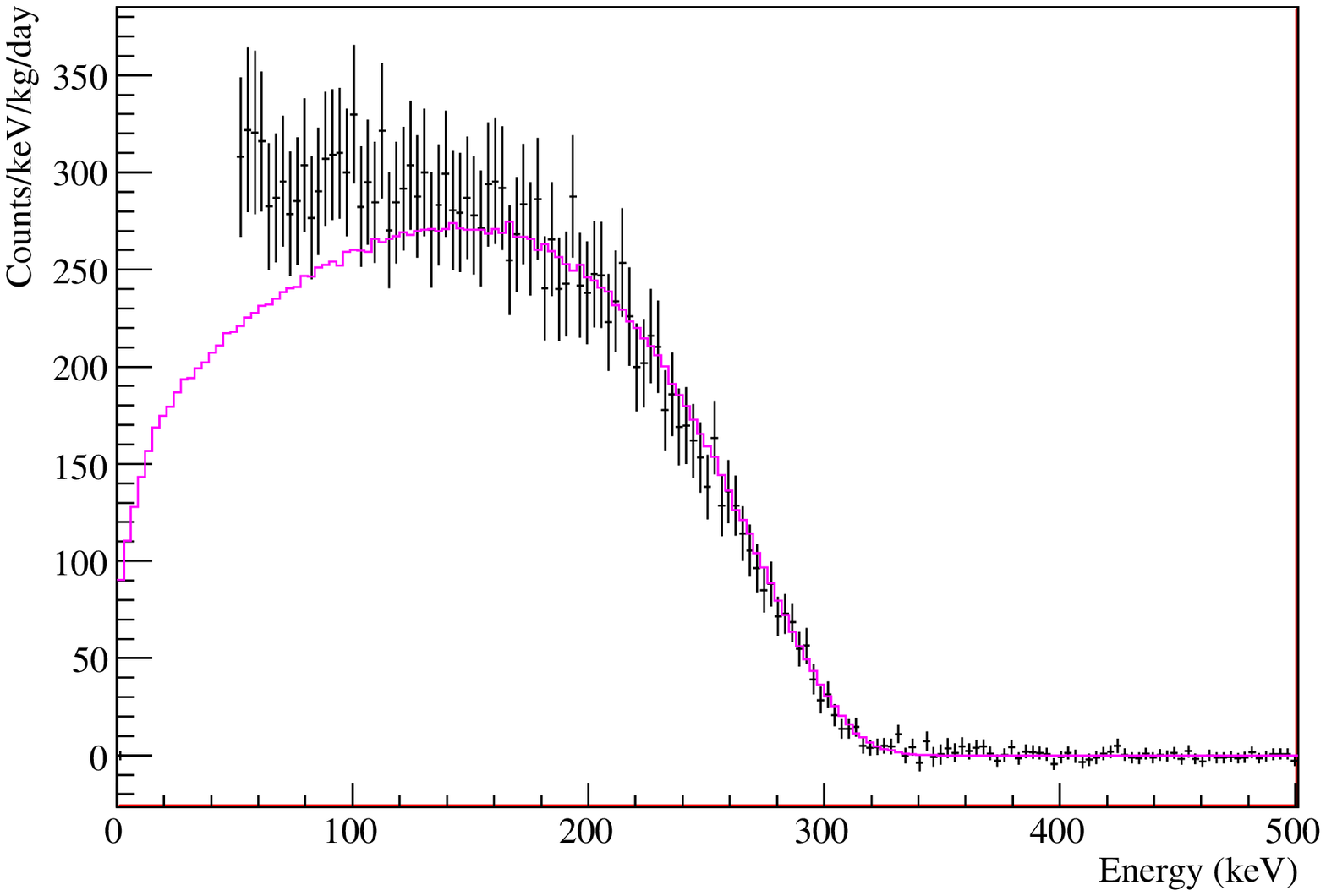}
}
\subfigure[Detector 15] 
{
\includegraphics[width=6cm,height=3cm]{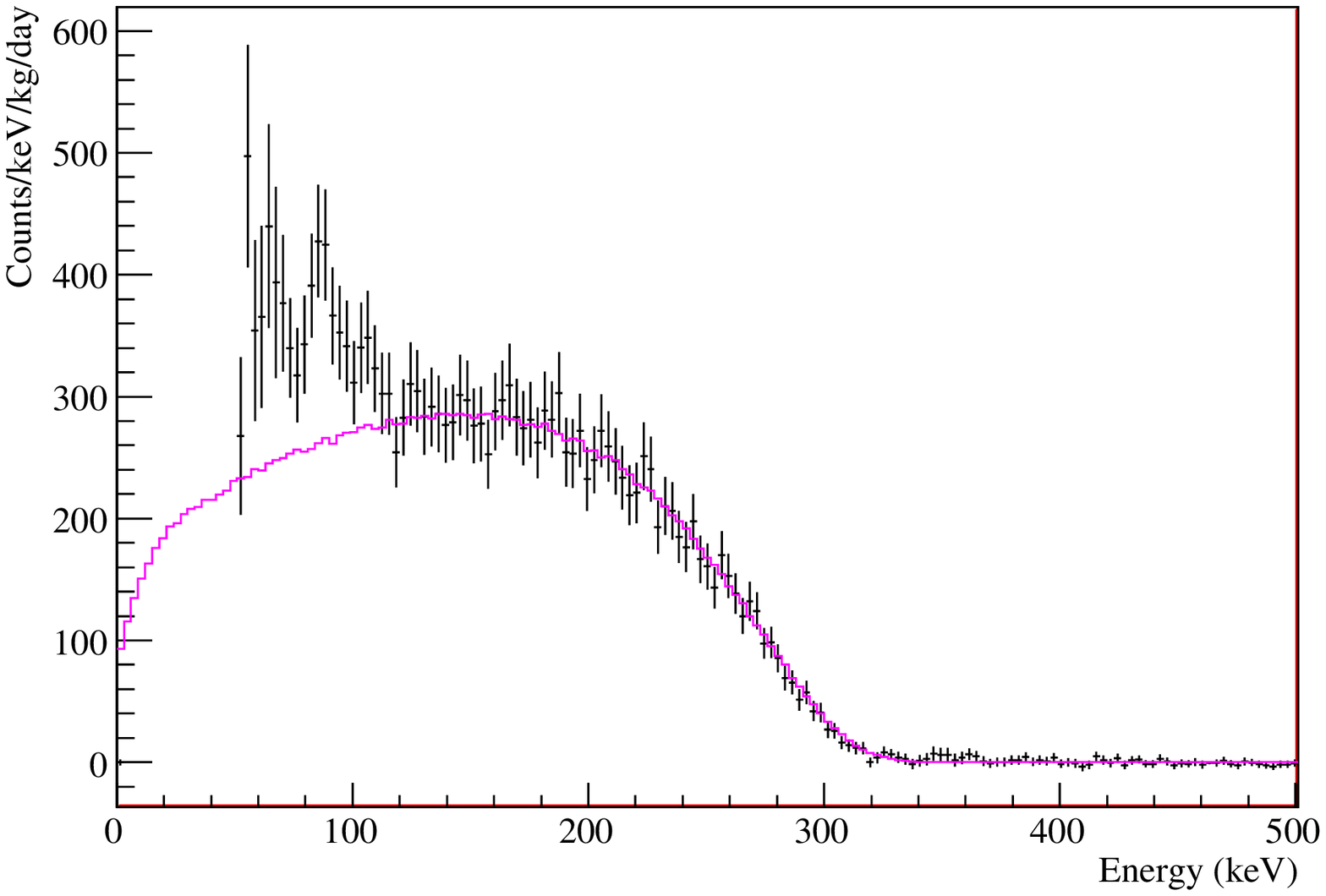}
}
\end{figure}

Each spectrum was fitted individually with a theoretical $^{113}$Cd
spectrum over the range 100 to 350 keV.  If the goodness-of-fit probability was low (less than 0.1) then the low energy threshold was raised to 150 keV.  The fit range, which passes the expected endpoint of the spectrum, is necessary to establish that the residual background is consistent with 0.  Above 150 keV we are confident that the background model accurately describes the data, and therefore the background subtraction is valid. \\
The Q value and spectral parameters ($c_1$, $c_2$ and $c_3$) should be the same for each detector since they describe the shape of the beta spectrum.  Previous estimates of the spectral shape parameters have been measured by other authors:
$c_1 = 1.01 \pm 0.01, c_2 = 1.48 \pm 0.05, c_3 = 0.68 \pm 0.21 $\cite{Danevich_1996}, $c_1 = 0.765 \pm 0.095, c_2 = 0.589 \pm 0.177, c_3 = 2.04 \pm 0.74 $\cite{Alessandrello_1994} and  $c_1 = 1.016 \pm 0.005, c_2 = 1.499 \pm 0.016, c_3 = 3.034 \pm 0.045$ \cite{Belli_2007}.\\
A first attempt at fitting all the spectra resulted with a best estimate of Q as 323 $\pm$ 0.4 keV with $c_1 = 0.97\pm0.03$, $c_2 =3.14\pm0.04$ and $c_3= -0.33\pm0.02$.  Only the value of $c_1$ measured here is consistent with these previous measurements. The reason being that $c_2$ and $c_3$ especially are highly dependent on the behaviour at low beta energies. The threshold in this work is too high to well constrain these parameters.\\
A recent publication by \cite{Belli_2007} gives a well measured beta spectrum of $^{113}$Cd down to 30 keV with the best uncertainties on the spectral parameters, $c_1 = 1.016 \pm 0.005, c_2 = 1.499 \pm 0.016, c_3 = 3.034 \pm 0.045$. The 11 spectra were re-fitted using these spectral shape values as inputs rather than as free-floating parameters. The uncertainties on each spectral parameter were used to ascertain the uncertainties on the half-life in the procedure described in \ref{fittingmethod}.  The fits are shown in Figure \ref{fig:sub}.\\
Table \ref{table:results2} shows the outputted fit parameters and their uncertainties. In this scenario the weighted mean Q value is 
$322.2\pm0.3(stat)$ keV.  Table \ref{table:halflives2} shows the resulting half-lives determined from this fit. Weighted means of these half-lives give a new estimate of $^{113}$Cd half-life as $(8.00 \pm 0.11(stat))\times10^{15}$ years.\\
\begin{table}[pth]
\caption{Table of Results}
{\footnotesize
\begin{tabular}{@{}ccccccc@{}}

{Detector} & {Q} &{Uncertainty} & {Normalisation}  &{Uncertainty} & {Fit range}&{Chi$^2$/NDF} \\[-1.5ex]
{} & { keV} & { keV } & { $\times 10^{-25}$} & { $\times 10^{-26}$} & {keV} & {}\tabularnewline
\hline
\hline
{ 2}&  { 322.27}&  { 1.24}&  { 3.18}&  { 1.1}&  { 100--350}&  { 70.8/83 }\\[-1.5ex]
{ 3}&  { 320.79}&  { 1.16}&  { 3.09}&  { 1.1}&  { 100--350}&  { 79.6/83 }\\[-1.5ex]
{ 5}&  { 321.06}&  { 1.08}&  { 3.35}&  { 1.1}&  { 100--350}&  { 41.7/83 }\\[-1.5ex]
{ 7}&  { 320.69}&  { 1.05}&  { 3.16}&  { 1.0}&  {100--350}&  { 57.1/83 }\\[-1.5ex]
{ 8}&  { 323.23}&  { 1.46}&  { 3.03}&  { 1.1}&  { 100--350}&  { 76.2/83 }\\[-1.5ex]
{ 9}&  { 323.47}&  { 1.17}&  { 3.05}&  { 1.1}&  { 100--350}&  { 65.1/83 }\\[-1.5ex]
{ 10}&  { 327.34}&  { 1.23}&  { 2.44}&  { 0.9}&  { 150--350}&  { 55.8/66 }\\[-1.5ex]
{ 11}&  { 322.03}&  { 1.25}&  { 2.74}&  { 1.1}&  { 150--350}&  { 58.3/66 }\\[-1.5ex]
{ 12}&  { 323.37}&  { 1.25}&  { 2.99}&  { 1.1}&  { 100--350}&  { 55.7/83 }\\[-1.5ex]
{ 14}&  { 321.05}&  { 1.07}&  { 3.12}&  { 1.0}&  { 100--350}&  { 45.9/83 }\\[-1.5ex]
{ 15}&  { 322.27}&  { 1.02}&  { 3.19}&  { 0.9}&  { 100--350}&  { 42.6/83 }\\[-1.5ex]
\end{tabular}\label{table:results2} }   
\end{table}

\begin{table}[pth]
\caption{Table of Results}
{\footnotesize
\begin{tabular}{@{}ccc@{}}
{Detector} & {Half-life (years)} & {Uncertainty (years)} \tabularnewline
\hline
\hline
{2} &{ $7.97\times 10^{15}$ } &{$0.35\times 10^{15}$  }\\[-1.5ex]
{3 } &{$8.23\times 10^{15 }$} &{$0.38\times 10^{15}$ }\\[-1.5ex]
{5} &{ $7.50\times 10^{15}$} &{$ 0.30\times 10^{15}$ }\\[-1.5ex]
{7} &{$ 8.01\times 10^{15}$} &{$0.32 \times 10^{15}$ }\\[-1.5ex]
{8 } &{$7.91\times 10^{15 }$} &{$0.45\times 10^{15}$ }\\[-1.5ex]
{9 } &{$7.79\times 10^{15 }$} &{$0.34\times 10^{15}$}\\[-1.5ex]
{10} &{$8.88\times 10^{15}$} &{$ 0.44\times 10^{15}$ }\\[-1.5ex]
{11} &{$8.97\times 10^{15}$} &{ $0.47\times 10^{15}$}\\[-1.5ex]
{12 } &{$7.97\times 10^{15 }$} &{$0.38\times 10^{15}$ }\\[-1.5ex]
{14 } &{$8.07\times 10^{15 }$} &{$0.32\times 10^{15}$ }\\[-1.5ex]
{15 } &{$ 7.66\times 10^{15}$} &{ $0.32\times 10^{15}$ }\\[-1.5ex]
\end{tabular}\label{table:halflives2} }   
\end{table}

\section{Discussion} \label{Discussion}
\begin{figure}
\centering{
\includegraphics[width=15cm]{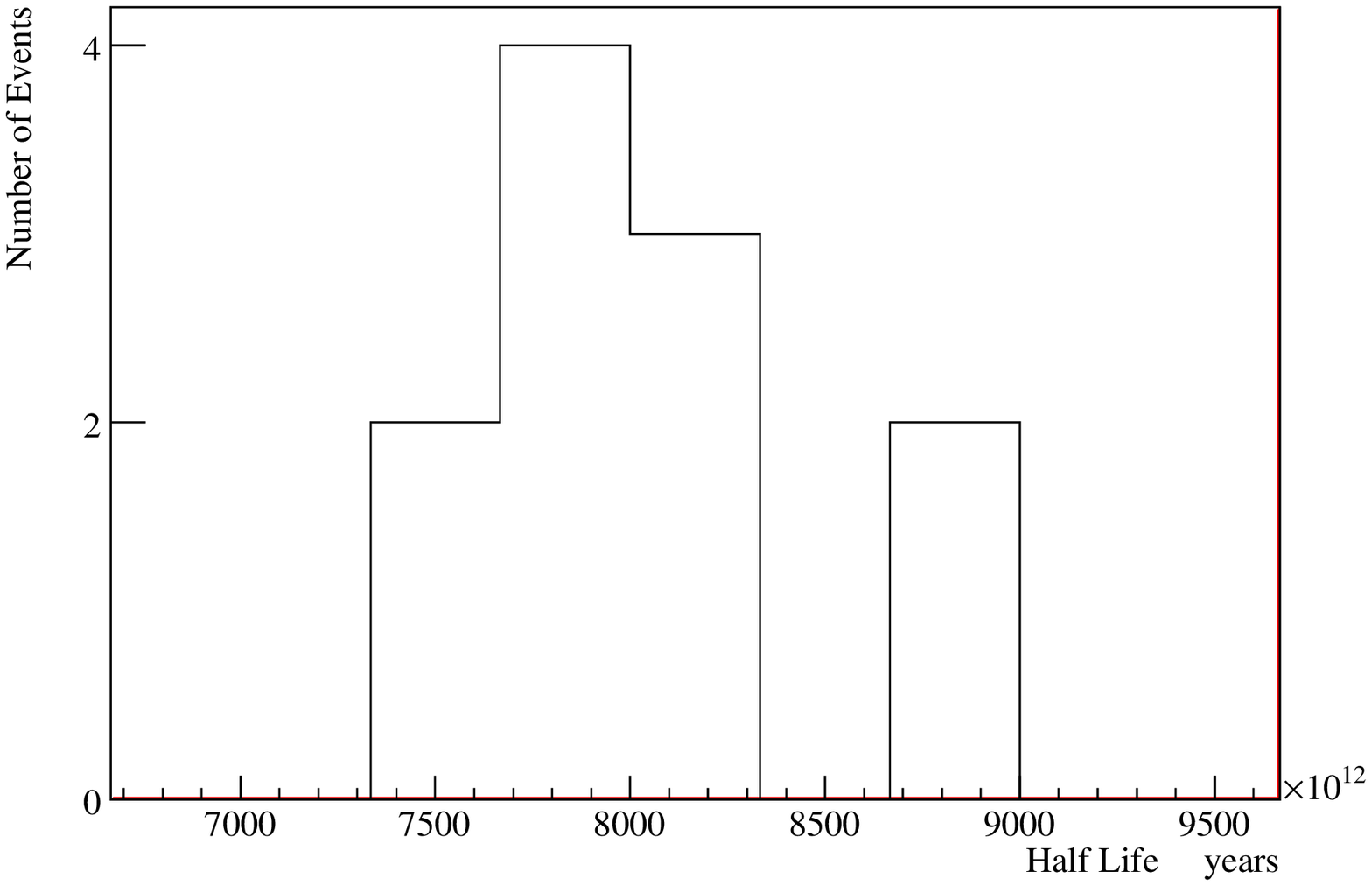}
\caption{Histogram of the determined half-lives for all 11 detectors.}
\label{fig:halflife_hist}}
\end{figure}

\begin{figure}
\centering{
\includegraphics[width=15cm]{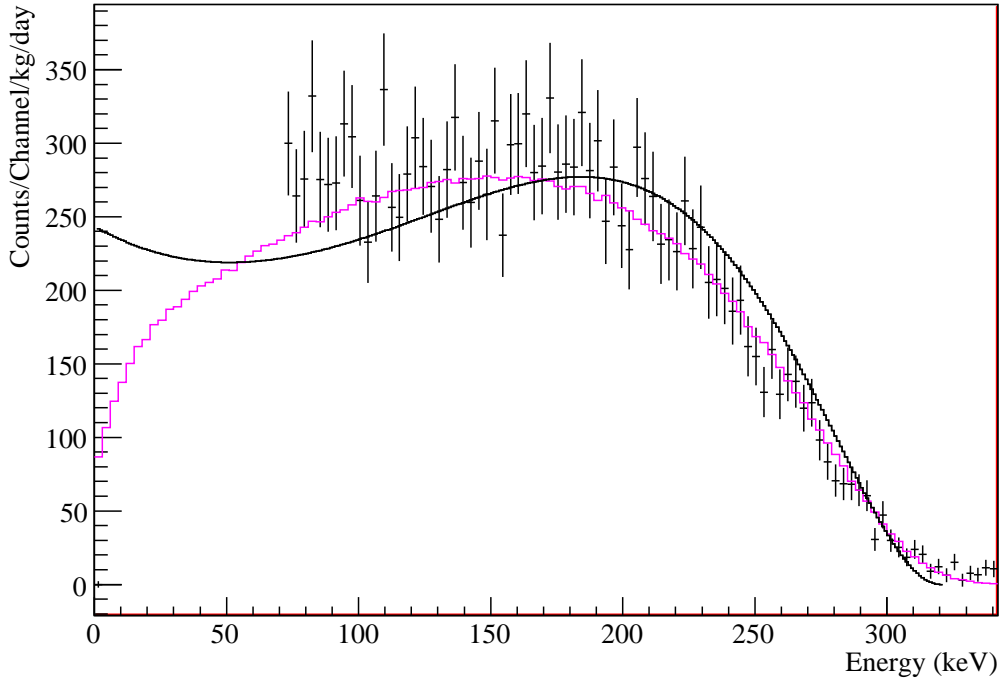}
\caption{$^{113}$Cd spectra from Detector 9 showing fit (histogram), and microscopic calculation of  \protect \cite{Jouni} (smooth line). Despite the fact that the major effect occurs in the low energy behaviour a discrepancy is already visible at the high energy end. This is true in all measured spectra.  The parameterisation of \cite{Belli_2007} is a good fit to the data above 100 keV.}
\label{fig:jouni}}
\end{figure}

We have made the most numerous measurements of 4-fold forbidden beta
decays in this case for $^{113}$Cd, using 11 independent CdZnTe
detectors. Each can have a unique (and independent) set of detector
parameters such as energy resolution, mass and $^{113}$Cd content etc.
This gives a clear cross-check on our estimate of total systematic
uncertainty.  Figure \ref{fig:halflife_hist} shows a histogram of the
half-lives determined for each detector.  The spread in these results
must be consistent (or at least smaller than) the combination of the
estimated total systematic uncertainty and the determined statistical
errors.  The total systematic uncertainty was estimated earlier in
Section \ref{Systematics} to be $\sim$3\%.  \\
The statistical uncertainties on each half-life are derived from the fits and are shown
in Table \ref{table:halflives2}.  The weighted mean uncertainty is $1.1 \times
10^{14}$ years which is 1.3\% of the weighted mean half-life. Assuming the systematic and statistical uncertainties are
independent, we would therefore expect the distribution of the half-lives
shown in Figure \ref{fig:halflife_hist} to have a standard deviation of $\sim$3.3\%.\\
As can be seen in Figure \ref{fig:halflife_hist}, all the half-life estimates cluster except for two results which are well separated from the rest of the data. These are from detectors 10 and 11, which both show significant low energy features and displayed bad goodness-of-fit probabilities when fitted with the low energy threshold of 100 keV.  If these two detectors are ignored then the remaining estimates give an overall half-life value of $(7.9 \pm 0.3)\times 10^{15}$ years. The total uncertainty found is therefore 4\% of the mean value. This uncertainty is larger than our estimates, 3.3\%, but not unreasonably so. It may indicate that we have underestimated one of our systematic errors.  One clear possibility to explain this discrepancy could be the assumed spread in the cadmium content since we have as yet no independent measurements of the content of these
detectors.\\
The estimate of the Q-value is dependent on the fitted function, and
assuming the used functional form to be correct, is found to be 322.2$\pm$0.3(stat)$\pm$0.9(sys) keV.  Interestingly the work of \cite{Belli_2007} fails to fit a feasible Q-value that is consistent with the Table of Isotopes value of 320 $\pm$ 3 keV \cite{wapstra}.\\
Encouraged by the results of \cite{Goessling_2005} for the first time
microscopic models were explored to predict the half-life and spectral
shapes of 4-fold forbidden non-unique beta decays \cite{Jouni}.  The
measured spectrum of one of the detectors, our fitted spectrum and
this theoretical calculation are shown in Figure \ref{fig:jouni}.  The
theoretical calculation assumed a Q-value of 320 keV.
The spectral shape we measure does not match well with the theoretical spectrum and instead agrees well with the 4-fold forbidden unique spectrum as shown by \cite{Belli_2007}. This is an interesting result and will hopefully motivate some further investigation as to why the fit for a unique transition, normally only linked to one nuclear matrix element fits better than the actual non-unique calculation.

\section{Conclusion}
\label{conclusion}      
CdZnTe semiconductor detectors were used to search for the rare decay of the 4-fold forbidden 
non-unique beta decay of $^{113}$Cd. For the first time a statistically relevant sample of half-life measurements have been obtained for such rare decays. 
We confirm that the parameterisation of \cite{Belli_2007} is a good fit to our data above 100 keV.  Using this parameterisation we obtain an estimate of the Q-value of the decay to be 322.2$\pm$0.3(stat)$\pm$0.9(sys) keV.  Assuming the fit to the data is correct, the resultant half-life is $(8.00 \pm 0.11(stat) \pm 0.24(sys))\times10^{15}$ years which is completely consistent with previous measurements.\\
Future work will improve upon the background levels experienced by the experiment.  Short test runs with four lower background detectors (with colourless passivation coatings) have already been made.  These new runs also featured a clean nitrogen flushing system which reduces significantly the radon level in the shielded setup.  Longer runs are required to improve on the current result.  Ideally a new long run with many colourless detectors would be done, allowing the same systematic cross-checking that has been done in this experiment.\\
As shown in Figure \ref{fig:sub} there are some unexplained background features below 100 keV that are present on most but not all of the detectors.  Future work will try to ascertain where this background comes from. \\
Applying mild cooling ($\sim$10$^{\circ}$C) to the detectors whilst
inside the shielded setup will also be explored. This will reduce the
surface leakage current dramatically and therefore lower the energy
threshold and improve the energy resolution \cite{dawson2008}.\\
Non-intrusive means of measuring the zinc content are already being
explored e.g. photoluminescence and X-ray fluorescence techniques.
However, these techniques must be applied to the crystals in use
either before bonding or at the end of the science run.

\subsection*{Acknowledgements}
We thank our COBRA colleagues for their support.  J Wilson acknowledges the support of the Leverhulme Trust. We are grateful to J. Suhonen for providing a data file with the expected beta spectrum based on microscopic calculations.

\end{document}